\documentstyle[twocolumn,11pt]{article}

\input epsf

\begin{document}

\date{}                                 

\title{\bf Web data modeling for integration in data warehouses}           


\author{Sami Miniaoui, J{\'e}r{\^o}me Darmont, Omar Boussaid \\
ERIC, Universit{\'e} Lumi{\`e}re Lyon 2 \\
5 avenue Pierre Mend{\`e}s-France \\
69676 Bron Cedex \\
France}

\maketitle                        
\thispagestyle{empty}


\noindent {\bf Abstract
      {\small\em In a data warehousing process, the data preparation phase is crucial. Mastering 
this phase allows substantial gains in terms of time and performance when performing a 
multidimensional analysis or using data mining algorithms. Furthermore, a data warehouse can 
require external data. The web is a prevalent data source in this context, but the data 
broadcasted on this medium are very heterogeneous. We propose in this paper a UML conceptual model 
for a complex object representing a superclass of any useful data source (databases, plain texts, 
HTML and XML documents, images, sounds, video clips...). The translation into a logical model is 
achieved with XML, which helps integrating all these diverse, heterogeneous data into a unified 
format, and whose schema definition provides first-rate metadata in our data warehousing context. 
Moreover, we benefit from XML's flexibility, extensibility and from the richness of the 
semi-structured data model, but we are still able to later map XML documents into a database if 
more structuring is needed.}
}

\vspace{0.5cm}



\section{Introduction}
\label{introduction}

In the context of e-commerce, analyzing the behavior of a customer, a product, or a company 
consists in monitoring one or several activities (commercial or medical pursuits, patent 
deposits, etc.). The objective of multidimensional analysis, particularly OLAP, is to analyze such 
activities under the form of numerical data. The information is summarized and can be presented as 
relevant information (i.e., knowledge), thus connecting OLAP to other analysis tools such as KDD 
({\em Knowledge Discovery in Databases}) techniques (namely, data mining), whose objectives are to 
understand and predict the behavior of one or several activities.

To be efficient in terms of quality and response time, analysis tools need their input data to be 
properly structured, acquired, and prepared in a previous step. These data are typically stored in 
databases aimed at decision support (such as data warehouses) that we call Decision Databases 
(DDBs). These databases can necessitate external data sources. For instance, a company willing to 
support competitive monitoring cannot merely analyze only data from its own production databases. 
The web is a prevalent data source in this context. However, the data broadcasted on this medium 
are very heterogeneous, which makes their conceptualization in a data warehousing framework 
difficult. Nonetheless, the concepts of data warehousing \cite{CHA97} remain valid in this 
approach. Measures, though not necessarily numerical, remain the indicators for analysis, and 
analysis is still performed following different perspectives represented by dimensions. Large data 
volumes and their dating are other arguments in favor of this approach \cite{KIM00}.

Our objective is to use the web as a full data source for DDBs, in a transparent way. This raises 
several issues:
\begin{itemize}
\item structuring multiform data from the web --- databases, plain texts, multimedia data (images, 
sounds, video clips...), semi-structured data (HTML, XML, or SGML documents) --- into a database;
\item integrating these data into the particular architecture of a data warehouse (fact tables, 
dimension tables, data marts);
\item devising evolution strategies for the warehouse when new data pop up;
\item physically reorganizing data depending on usage to improve query performance.
\end{itemize}

The aim of this paper is to address the first issue. We propose a unified UML model for a complex 
object representing a superclass of the multiform data we need to integrate in a DDB. Our objective 
is not only to store data, but also to truly prepare them for analysis. This is not a mere ETL 
({\em Extracting, Transforming, and Loading}) task, which would only render data names and domains 
consistent.

We elected to translate our UML conceptual model into an XML \cite{BRA00} logical model, for several reasons. 
First, XML encapsulates both data and their schema, either implicitly or in a DTD. This representation
is also found in data warehouses, which store both data and metadata that describe the data. Hence, 
XML is particularly adapted for our purposes. Moreover, we benefit from the flexibility, the 
extensibility and the richness of the semi-structured data model. And since XML documents can easily 
be mapped into a conventional (e.g., relational) database \cite{AND00}, we can also take advantage of 
well-structured data and query processing efficiency, 
if necessary when we move down to the physical model. Furthermore, XML-based 
databases like Lore \cite{MCH97} are quickly expanding. Hence, we get the best of both worlds (the 
structured and the semi-structured) by adopting the XML format.
 
The remainder of this paper is organized as follows. Section~\ref{related_work} establishes a short 
state of the art regarding XML mapping and federated, multimedia, and text databases. 
Section~\ref{uml} presents our unified model for multiform data. Section~\ref{xml} outlines how this 
conceptual model is translated into a logical, XML model. We finally conclude the paper and discuss 
future research issues in Section~\ref{conclusion}.

\section{Related work}
\label{related_work}

\subsection{Data integration}

We identified two main ways to integrate heterogeneous data into a data warehouse. The first approach 
relates to federated databases, that are distributed and heterogeneous databases constituted from data
sources of various natures: HTML or XML documents, databases, and so on, and providing users with an 
integrated view of the data \cite{BUS99, BER01}. The casual architecture for a federated database is 
layered in three levels:
\begin{itemize}
\item {\em presentation:} components allowing to formulate queries in the federated database language;
\item {\em mediation:} mediators in charge of collecting queries issued by users in the presentation
components and translating them in the proper language of each data source;
\item {\em adaptation:} components allowing communication between data sources and mediators.
\end{itemize}

The second possible approach consists in capturing the common characteristics of the different data 
types we need to integrate. In order to propose a unified data model, we took interest in how data is 
structured, stored, and indexed in textual and multimedia databases. Indexing strategies in textual 
databases include reversed lists of significant terms with their frequency of appearance in each 
document, signatures obtained by hashing keywords, and relative frequency matrix of words present in 
a set of documents \cite{GAR99a}. Multimedia databases may adopt the following characteristics to 
index images: signatures derived from (manually captured) keywords describing the image, color, 
texture or brightness distributions, etc. \cite{RAK95}

\subsection{XML mapping}

As XML emerged as a data exchange standard language, its storage in databases became a research issue.
Several approaches have been adopted to map an XML document into a data-base. \cite{BOO99, KAP00} 
propose algorithms that exploit a UML schema to map a DTD into a relational schema. Another approach 
consists in representing an XML document as a labeled, oriented graph where vertices are data types, 
edges are classes or objects, and leaves are data \cite{GAR99b}. In these two approaches, a relational
or object-relational database system was used to store the XML documents, but XML-native DBMSs, such 
as LORE \cite{MCH97}, also exist.

\section{Unified conceptual model}
\label{uml}

\begin{figure*}[tb]   
\begin{center}
\epsfxsize=16cm
\centerline{\epsffile{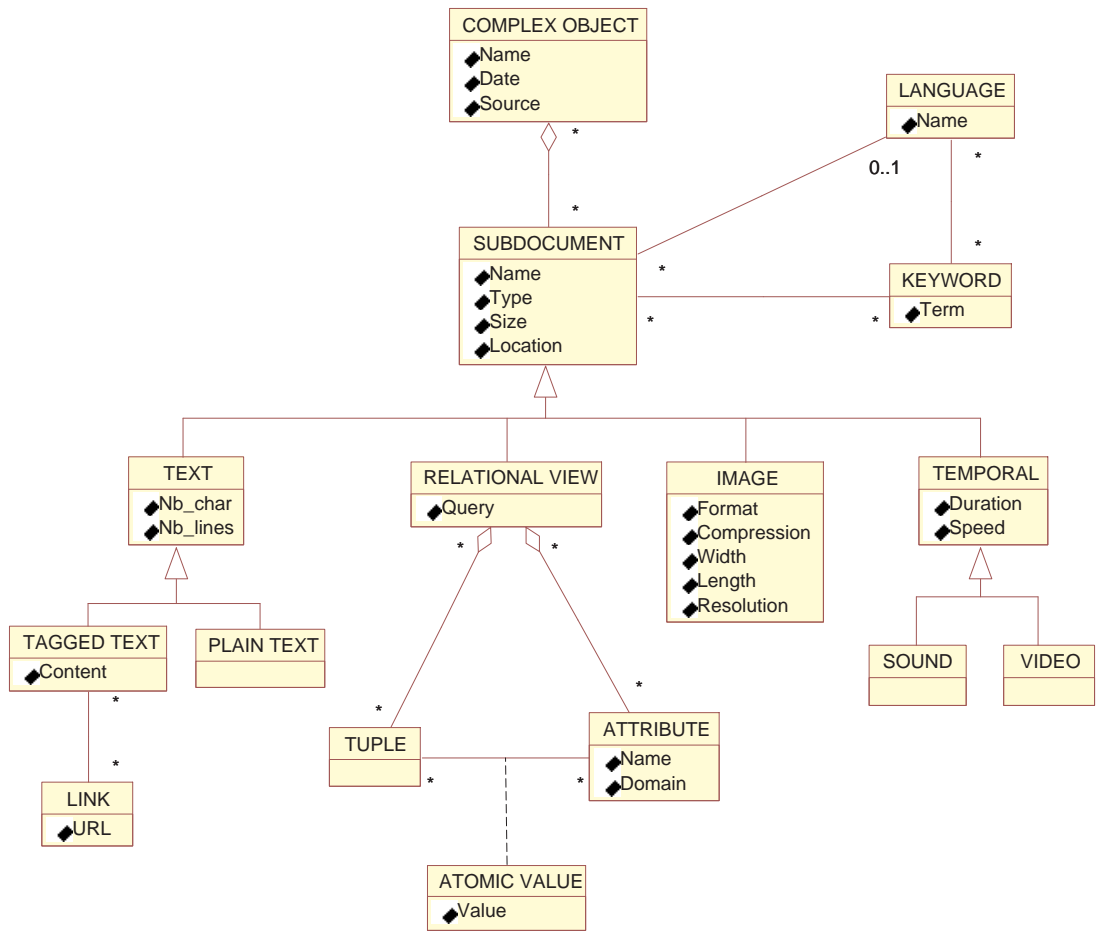}}
\caption{Multiform data model}
\label{fig1}
\end{center}
\end{figure*}

The data types we consider (text, multimedia documents, relational views from databases) for 
integration in a data warehouse all bear characteristics that can be used for indexing. The UML class 
diagram shown in Figure~\ref{fig1} represents a complex object generalizing all these data types. 
Note that our goal here is to propose a general data structure: the list of attributes for each class 
in this diagram is willingly not exhaustive.

A complex object is characterized by its name and its source. The date attribute introduces the 
notion of successive versions and dating that is crucial in data warehouses. Each complex object is 
composed of several subdocuments. Each subdocument is identified by its name, its type, its size, and 
its location (i.e., its physical address). The document type (text, image, etc.) will be helpful 
later, when selecting an appropriate analysis tool (text mining tools are different from standard 
data mining tools, for instance). The language class is important for text mining and information 
retrieval purposes, since it characterizes both documents and keywords.

Eventually, keywords represent a semantic representation of a document. They are metadata describing 
the object to integrate (medical image, press article...) or its content. Keywords are essential in 
the indexing process that helps guaranteeing good performances at data retrieval time. Note that we 
consider only logical indexing here, and not physical issues rose by very large amounts of data, 
which are still quite open as far as we know. Keywords are typically manually captured, but it would 
be very interesting to mine them automatically with text mining, image mining, or XML mining (tag 
detection) techniques, for instance.

All the following classes are subclasses of the subdocument class. They represent the basic data 
types and/or documents we want to integrate. Text documents are subdivided into plain texts and 
tagged texts (namely HTML, XML, or SGML documents). 
Tagged text are further associated to a certain number of links. Since a web 
page may point to external data (other pages, images, multimedia data, files...), 
those links help relating these data to their refering page.
 
Relational views are actually extractions from any type of database (relational, object, 
object-relational --- we suppose a view can be extracted whatever the data model) that will be 
materialized in the data warehouse. A relational view is a set of attributes (columns, classically 
characterized by their name and their domain) and a set of tuples (rows). At the intersection of 
tuples and attributes is a data value. In our model, these values appear as ordinal, but
in practice they can be texts or BLOBs containing multimedia data. The query that 
helped building the view is also stored. Depending on the context, all the data can 
be stored, only the query and the intention (attribute definitions), or everything. For instance, it might be inadequate to duplicate
huge amounts of data, especially if the data source is not regularly updated. On
the other hand, if successive snapshots of an evolving view are needed, data will
have to be stored.

Images may bear two types of attributes: some that are usually found in the image file header 
(format, compression rate, size in pixels, resolution), and some that need to be extracted by 
program, such as color or texture distributions.

Eventually, sounds and video clips are part of a same class because they share temporal attributes 
that are absent from the other types of data we consider. As far as we know, these types of data are 
not currently analyzed by mining algorithms, but they do contain knowledge. This is why we take them 
into account here (though in little detail), anticipating advances in "multimedia mining" techniques.

\section{XML-based logical model}
\label{xml}

In a data warehouse, data come with metadata that describe their origin, the rules for transformations
they may have undergone, and information regarding their usage \cite{WU97}. Identically, multiform 
data modeled as complex objects may be viewed as {\em documents}. It is then necessary to consider 
two kinds of data: data themselves and sets of information pieces, such as descriptive data that 
allow their identification, or status data describing their semantics.

The use of XML as an implementation tool for multiform data (i.e., complex objects) viewed as 
documents seems natural. This language indeed helps representing both the description and the content 
of any document. Within a modeling process, the translation of UML classes representing multiform 
data into XML constitutes a logical formalization phase. The obtained logical model can be as well 
mapped in a relational or object-relational database as stored in a native XML database.

\subsection{XML DTD}

When integrating multiform data, we adopt a classical information system modeling process: first 
devise a conceptual model, and then translate it into a logical model. The UML class diagram from 
Section~\ref{uml} is our conceptual model. We consider XML as a fine candidate for logical modeling.

The UML model can indeed be directly translated into an XML DTD ({\em Document Definition Type}),
as shown in Figure~\ref{fig2}. However, we applied minor shortcuts not to overload the DTD. 
Since the {\em LANGUAGE}, {\em KEYWORD}, {\em LINK}, and {\em VALUE} classes only bear one attribute each, we mapped them to
single XML elements, rather than having them be composed of another, single element. For instance, 
the {\em LANGUAGE} class became the {\em LANGUAGE} element, but this element is not further composed
of the {\em Name} element. Eventually, since the {\em ATTRIBUTE} and the {\em TUPLE} elements share the
same sub-element "attribute name", we labeled it {\em ATT\_NAME} in the {\em ATTRIBUTE} element and 
{\em ATT\_NAME\_REF} (reference to an attribute name) in the {\em TUPLE} element to avoid any confusion
or processing problem.

\begin{figure*}[tbp]   
\begin{center}
\begin{tabular}{|p{16cm}|}
\hline
{\small {\tt 
<!ELEMENT COMPLEX\_OBJECT (OBJ\_NAME, DATE, SOURCE, SUBDOCUMENT+)> \newline
\hspace*{0.5cm} <!ELEMENT  OBJ\_NAME       PCDATA \#REQUIRED>\newline
\hspace*{0.5cm} <!ELEMENT  DATE         PCDATA \#REQUIRED>\newline
\hspace*{0.5cm} <!ELEMENT  SOURCE   PCDATA \#REQUIRED>\newline
\hspace*{0.5cm} <!ELEMENT  SUBDOCUMENT (DOC\_NAME, TYPE, SIZE, LOCATION, LANGUAGE?, KEYWORD*, (TEXT | RELATIONAL\_VIEW | IMAGE | TEMPORAL))>\newline
\hspace*{1cm} <!ELEMENT  DOC\_NAME     PCDATA \#REQUIRED>\newline
\hspace*{1cm} <!ELEMENT  TYPE                PCDATA \#REQUIRED>\newline
\hspace*{1cm} <!ELEMENT  SIZE                 PCDATA \#REQUIRED>\newline
\hspace*{1cm} <!ELEMENT  LOCATION     PCDATA \#REQUIRED>\newline
\hspace*{1cm} <!ELEMENT  LANGUAGE  PCDATA \#REQUIRED>\newline
\hspace*{1cm} <!ELEMENT  KEYWORD  PCDATA \#REQUIRED>\newline  
\hspace*{1cm} <!ELEMENT  TEXT  (NB\_CHAR, NB\_LINES, (PLAIN\_TEXT | TAGGED\_TEXT))>\newline
\hspace*{1.5cm} <!ELEMENT   NB\_CHAR          PCDATA \#IMPLIED>\newline
\hspace*{1.5cm} <!ELEMENT  NB\_LINES           PCDATA \#IMPLIED>  \newline
\hspace*{1.5cm} <!ELEMENT  PLAIN\_TEXT       PCDATA \#REQUIRED>\newline
\hspace*{1.5cm} <!ELEMENT  TAGGED\_TEXT   (CONTENT, LINK*)>\newline
\hspace*{2cm} <!ELEMENT  CONTENT       PCDATA \#REQUIRED>\newline
\hspace*{2cm} <!ELEMENT  LINK       PCDATA \#REQUIRED>\newline
\hspace*{1cm} <!ELEMENT  RELATIONAL\_VIEW (QUERY?, ATTRIBUTE+, TUPLE*)>\newline
\hspace*{1.5cm} <!ELEMENT  QUERY       PCDATA \#REQUIRED>\newline
\hspace*{1.5cm} <!ELEMENT  ATTRIBUTE   (ATT\_NAME, DOMAIN)>\newline
\hspace*{2cm} <!ELEMENT  ATT\_NAME  PCDATA \#REQUIRED>\newline
\hspace*{2cm} <!ELEMENT  DOMAIN        PCDATA \#REQUIRED>\newline
\hspace*{1.5cm} <!ELEMENT  TUPLE   (ATT\_NAME\_REF, VALUE)+>\newline
\hspace*{2cm} <!ELEMENT  ATT\_NAME\_REF  PCDATA \#REQUIRED>\newline
\hspace*{2cm} <!ELEMENT  VALUE           PCDATA \#IMPLIED>    \newline
\hspace*{1cm} <!ELEMENT  IMAGE (COMPRESSION, FORMAT, RESOLUTION, LENGTH, WIDTH)>\newline
\hspace*{1.5cm} <!ELEMENT  COMPRESSION  PCDATA \#IMPLIED> \newline
\hspace*{1.5cm} <!ELEMENT  FORMAT             PCDATA \#IMPLIED>\newline
\hspace*{1.5cm} <!ELEMENT  RESOLUTION     PCDATA \#IMPLIED>\newline
\hspace*{1.5cm} <!ELEMENT  LENGTH              PCDATA \#IMPLIED>\newline
\hspace*{1.5cm} <!ELEMENT  WIDTH                PCDATA \#IMPLIED>\newline
\hspace*{1cm} <!ELEMENT  TEMPORAL (DURATION, SPEED, (SOUND | VIDEO))>\newline
\hspace*{1.5cm} <!ELEMENT  DURATION          PCDATA \#IMPLIED > \newline
\hspace*{1.5cm} <!ELEMENT  SPEED                  PCDATA \#IMPLIED>\newline
\hspace*{1.5cm} <!ELEMENT  SOUND                 PCDATA \#IMPLIED>\newline
\hspace*{1.5cm} <!ELEMENT  VIDEO                   PCDATA \#IMPLIED>
}}\\
\hline
\end{tabular}
\caption{XML DTD}
\label{fig2}
\end{center}
\end{figure*}

\subsection{Transformation algorithm}

We are currently in the process of developing a prototype capable of taking as input any data source 
from the web, fitting it in our model, and producing an XML document. The general algorithm for 
integrating multiform data in our unified model is provided in Figure~\ref{fig3}. 
It exploits the DTD from Figure~\ref{fig2}. The actual application is written in Java.

\begin{figure*}[tbp]   
\begin{center}
\begin{tabular}{|p{16cm}|}
\hline
{\small {\tt 
	{\bf // Initialization}\newline
	Read DTD line \newline
	Stack root element \newline
	{\bf // Main loop}\newline
	While stack not empty do \newline
\hspace*{0.5cm}	Unstack element \newline
\hspace*{0.5cm} {\bf // Positioning on the current element description}\newline
\hspace*{0.5cm}	While element not found in the DTD and not EOF(DTD) do \newline
\hspace*{1cm}			Read DTD line \newline
\hspace*{0.5cm}	End while \newline
\hspace*{0.5cm} 	If element was found then \newline
\hspace*{1cm} 		For each value of the element do {\bf // For elements with + or * cardinality}\newline
\hspace*{1.5cm} 			If element is atomic then \newline
\hspace*{2cm} 					Write elementBeginTag, elementValue, elementEndTag \newline
\hspace*{1.5cm} 			Else {\bf // Composite element} \newline
\hspace*{2cm}					Write elementBeginTag \newline	
\hspace*{2cm} 					Stack element {\bf // Necessary to later write end tag}\newline 
\hspace*{2cm} 					For each sub-element (in reverse order) do \newline
\hspace*{2.5cm} 					If sub-element does not belong to a selection then \newline
\hspace*{2.5cm}						{\bf // If element not in a list of the form (PLAIN\_TEXT $|$ TAGGED\_TEXT)}\newline
\hspace*{3cm} 						Stack sub-element \newline
\hspace*{2.5cm} 					Else \newline
\hspace*{3cm} 						If sub-element was selected then  \newline
\hspace*{3cm}						{\bf // If the DTD document type matches the actual document type}
\hspace*{3.5cm} 							Stack sub-element \newline
\hspace*{3cm} 						End if \newline
\hspace*{2.5cm} 					End if \newline
\hspace*{2cm} 					End for \newline
\hspace*{1.5cm} 			End if \newline
\hspace*{1cm} 		End for \newline
\hspace*{0.5cm} 	Else \newline
\hspace*{1cm} 		Write elementEndTag {\bf // Close composite elements}\newline
\hspace*{0.5cm} 	End if \newline
	End while \newline
}}
\\
\hline
\end{tabular}
\caption{Multiform data integration algorithm}
\label{fig3}
\end{center}
\end{figure*}

The principle of this algorithm is to browse the DTD, fetching the elements it describes,
and to write them into the XML document, along with the associated values extracted
from the original data, on the fly. Note that, when reading a DTD line, the current element we refer to
is the one which is being described, e.g., {\em TEXT} in the {\tt <!ELEMENT  TEXT  (NB\_CHAR, NB\_LINES, (PLAIN\_TEXT | TAGGED\_TEXT))>}
DTD line. We also suppose that sub-elements are defined in the same order they are declared in their
parent element.
Missing values are currently treated by inserting
an empty element, but strategies could be devised to solve this problem, either
by prompting the user or automatically.

At this point, our prototype is able to process text, image, and multimedia data. 
Figures~\ref{fig4} and \ref{fig5}
illustrate how multiform data (namely, an SGML tagged text and an image) are transformed 
using our approach.

\begin{figure*}[tb]   
\begin{center}
\begin{tabular}{|p{8.1cm}|p{8cm}|}
\hline
{\small {\bf SGML document}} & {\small {\bf XML model}} \\
\hline \hline
{\small {\tt <!DOCTYPE lewis SYSTEM "lewis.dtd">\newline
<REUTERS TOPICS="YES" LEWISSPLIT="TRAIN" CGISPLIT="TRAINING-SET" OLDID="12509" NEWID="326">\newline
<DATE>2-MAR-1987 06:41:06.17</DATE>\newline
<PLACES><D>france</D></PLACES>\newline
<COMPANIES>SNCF</COMPANIES>\newline
<TEXT>\newline
<TITLE>SNCF ISSUING THREE BILLION FRANC DOMESTIC BOND</TITLE>\newline
<DATELINE>PARIS, March 2</DATELINE>\newline
<BODY>The French state railway company, the Ste Nationale des Chemins de Fer Francaise (SNCF), is issuing a 
three billion French franc domestic bond in two tranches, the bond issuing committee said. Details of 
the issue will be announced later and it will be listed in the Official Bulletin (BALO) of March 9.
\newline The issue will be co-led by Banque Nationale de Paris, Caisse Nationale de Credit Agricole 
and the Societe Marseillaise de Credit.
\newline REUTER</BODY> </TEXT>\newline
</REUTERS>}}
&
{\small {\tt <?XML version=1.0?>\newline
<!DOCTYPE MultiformData SYSTEM "mlfd.dtd">\newline
<COMPLEX\_OBJECT>\newline
\hspace*{0.25cm}<NAME>Reuters Press Release</NAME>\newline
\hspace*{0.25cm}<DATE>May 15, 2001</DATE>\newline
\hspace*{0.25cm}<SOURCE>Reuters</SOURCE>\newline
\hspace*{0.25cm}<SUBDOCUMENT>\newline
\hspace*{0.5cm}<NAME>SGMLdoc</NAME>\newline
\hspace*{0.5cm}<TYPE>SGML</TYPE>\newline
\hspace*{0.5cm}<SIZE>820 Bytes</SIZE>\newline
\hspace*{0.5cm}<LOCATION>SGMLfile.sgml</LOCATION>\newline
\hspace*{0.5cm}<LANGUAGE>English</LANGUAGE>\newline
\hspace*{0.5cm}<KEYWORD>France</KEYWORD>\newline
\hspace*{0.5cm}<KEYWORD>SNCF</KEYWORD>\newline
\hspace*{0.5cm}<TEXT>\newline
\hspace*{0.75cm}<NB\_CHAR>790</NB\_CHAR>\newline
\hspace*{0.75cm}<NB\_LINES>12</NB\_LINES>\newline
\hspace*{0.75cm}<TAGGED\_TEXT>\newline
\hspace*{1cm}<CONTENT>{\em The document could be reproduced here as a CDATA.}</CONTENT>\newline
\hspace*{0.75cm}</TAGGED\_TEXT>\newline
\hspace*{0.5cm}</TEXT>\newline
\hspace*{0.25cm}</SUBDOCUMENT>\newline
</COMPLEX\_OBJECT>}}\\
\hline
\end{tabular}
\caption{Sample logical model for a tagged text}
\label{fig4}
\end{center}
\end{figure*}

\begin{figure*}[tbp]   
\begin{center}
\begin{tabular}{|p{8cm}|p{8cm}|}
\hline
{\small {\bf Image}} & {\small {\bf XML model}} \\
\hline\hline
\epsfxsize=7cm
\centerline{\epsffile{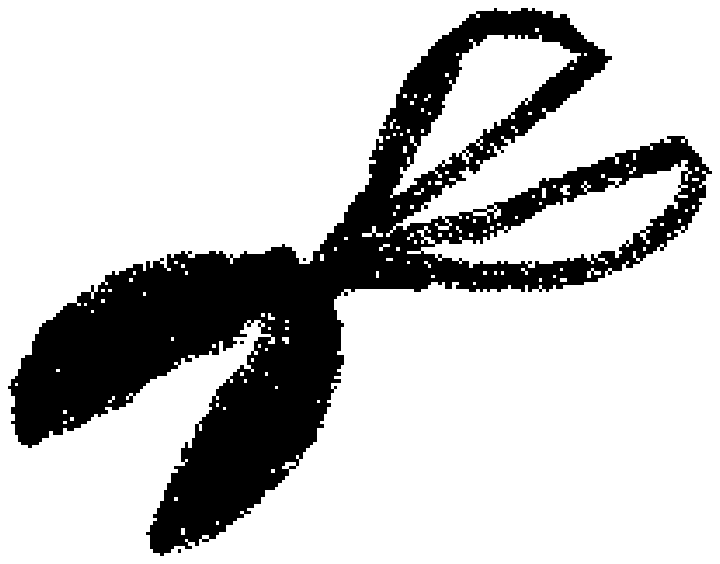}} 
{\small User-prompted keywords: scissors, black, white}
&
{\small {\tt <?XML version=1.0?>\newline
<!DOCTYPE MultiformData SYSTEM "mlfd.dtd">\newline
<COMPLEX\_OBJECT>\newline
\hspace*{0.25cm}<NAME>Sample image</NAME>\newline
\hspace*{0.25cm}<DATE>2001-06-15</DATE>\newline
\hspace*{0.25cm}<SOURCE>Local</SOURCE>\newline
\hspace*{0.25cm}<SUBDOCUMENT>\newline
\hspace*{0.5cm}<NAME>Scissors</NAME>\newline
\hspace*{0.5cm}<TYPE>Image</TYPE>\newline
\hspace*{0.5cm}<SIZE>24694 Bytes</SIZE>\newline
\hspace*{0.5cm}<LOCATION>scissors.bmp</LOCATION>\newline
\hspace*{0.5cm}<KEYWORD>scissors</KEYWORD>\newline
\hspace*{0.5cm}<KEYWORD>black</KEYWORD>\newline
\hspace*{0.5cm}<KEYWORD>white</KEYWORD>\newline
\hspace*{0.5cm}<IMAGE>\newline
\hspace*{0.75cm}<FORMAT>Bitmap</FORMAT>\newline
\hspace*{0.75cm}<COMPRESSION>None</COMPRESSION>\newline
\hspace*{0.75cm}<WIDTH>256</WIDTH>\newline
\hspace*{0.75cm}<LENGTH>192</LENGTH>\newline
\hspace*{0.75cm}<RESOLUTION>100 dpi</RESOLUTION>\newline
\hspace*{0.5cm}</IMAGE>\newline
\hspace*{0.25cm}</SUBDOCUMENT>\newline
</COMPLEX\_OBJECT>}}\\
\hline
\end{tabular}
\caption{Sample logical model for an image}
\label{fig5}
\end{center}
\end{figure*}

\section{Conclusion and future issues}
\label{conclusion}

We presented in this paper a UML model for a complex object that generalizes the different multiform 
data that can be found on the web and that are interesting to integrate in a data warehouse as 
external data sources. Our model allow the unification of these different data into a single 
framework, for purposes of storage and, maybe more importantly, preparation for analysis. Data must 
indeed be properly "formatted" before OLAP or data mining techniques can apply to them.

Our UML conceptual model is then directly translated into an XML DTD and instantiated into an XML 
document, which we both view as part of a logical model in our (classical) modeling process. XML is 
the format of choice for both storing and describing the data. The DTD indeed represents the metadata.
It is also very interesting because of its flexibility and extensibility, while allowing straight 
mapping into a more conventional database if strong structuring and retrieval efficiency
are needed for analysis purposes.

Since this work is only at its premises, perspectives are numerous. The first, immediate task to make 
this work concrete is completing our prototype by making it capable of taking as input not only texts
and multimedia data, but all the web data sources we identified in Figure~\ref{fig1}. 
The problem of missing values also needs to be addressed in more details since we could reuse existing,
automatic techniques.

In our next step, we will have to
integrate the documents produced by our application into a data 
warehouse. We suppose that mapping the XML document into a relational or object-relational database 
will be straightforward using the techniques presented in \cite{AND00}. 

Our XML modeling could also be improved by taking advantage of the features proposed in XML Schema
\cite{FAL01} that are not supported by DTDs, such as typing and inheritance.
In other respects, our XML formalization
may also be considered as first-level logical modeling. A multidimensional 
representation with dimensions and facts would make up the second level and allow the warehousing of 
multiform data. This second modeling level has not been discussed in this paper, but it is one of our
main goals for the integration of web data in a data warehouse.

Next, we do not envisage data mining as a front-end tool only. We believe integrating multi-form data 
in a data warehouse requires more than a simple ETL: it requires intelligence. We plan to include 
intelligence into the very preparation of data by using data mining techniques to extract information 
that is useful for storage (indexing), multidimensional analysis or data mining itself. For instance, 
we could automatically build a list of pertinent keywords for a given document, whether it is a text 
or a multimedia file, or extract such characteristics as color, texture, and brightness distributions 
from an image. The pre-processing of these characteristics would save time for ulterior analysis. 
Clustering techniques could also be used as a new type of aggregation of multiform data in a context 
of multidimensional analysis.

Eventually, monitoring data usage could help physically reorganizing the data to enhance the data 
warehouse's response time. For instance, new clusters could be devised to improve the performance of 
multidimensional queries.


\end{document}